\documentclass{article}
\usepackage[utf8]{inputenc}
\usepackage{latexsym,float}
\usepackage{authblk}
\usepackage{epsfig}
\usepackage{epstopdf}
\usepackage{amsmath,hyperref}
\usepackage{amssymb}
\usepackage{graphicx}
\usepackage{eepic}
\usepackage{bbm}
\usepackage{textcomp}
\usepackage{xcolor}
\usepackage[T1]{fontenc}

\title{Creation of wormholes during the cosmological bounce }

\begin{document}
\author[1]{Petar Pavlović \thanks{petar.pavlovic@kozmologija.org}}
\author[2,1]{Marko Sossich \thanks{marko.sossich@fer.hr}}
\affil[1]{\textit{\small{Institute for Cosmology and Philosophy of Nature, Trg svetog Florijana 16, Križevci, Croatia}}}
\affil[2]{\textit{\small{University of Zagreb, Faculty of Electrical Engineering and Computing, Department of Physics,
	Unska 3, 10 000 Zagreb, Croatia}}}

\maketitle

\begin{abstract}
In this work we demonstrate that wormholes can in principle be naturally
created during the cosmological bounce without the need for the exotic 
matter or any kind of additional modifications of the gravitational sector,
apart from the one enabling the cosmological bounce. This result is general
and does not depend on the details of the modifications of gravitational
equations needed to support the bounce. To study the possible existence of 
wormholes around the cosmological bounce we introduce general modifications
of Einstein's field equations need to support the bouncing solutions. In 
this regime we show that it is possible to construct a cosmological wormhole
solution supported by matter, radiation and vacuum energy, satisfying the
Weak Energy Condition (WEC), which asymptotically approaches the 
Friedmann-Lema\^{i}tre-Robertson-Walker (FLRW) metric.
However, at a specific cosmological time, which depends on the parameters 
of the bouncing cosmological model, the WEC describing the matter needed 
to support such wormholes is spontaneously violated. 
This means that such wormholes could potentially 
exist in large numbers  during some period around the bounce, 
significantly changing the causal structure of space-time, and then 
vanish afterwards.

\end{abstract}

\section{Introduction}
Our current understanding of the cosmological evolution is based on 
Einstein's general theory of relativity, which is one of the most 
successful physical theories. General theory of relativity was so far 
verified by various types of experiments -- from the light deflection 
and the perihelion advance of Mercury to the recent detection of 
gravitational waves \cite{prvi, drugi, treci, cetvrti}. One of the 
consequences of this theory is the necessary existence of singularities 
if the space-time is respecting some usual causal properties and if the 
matter is respecting the usual energy conditions (there are different 
variants of this result including the Strong, Null or Weak Energy 
Condition), as proven by the singularity theorems of
Hawking \cite{haw1, haw2, haw3}. One consequence of this result is the 
necessary existence of singularity in the evolution of our 
Universe, called the big bang singularity, if general theory of 
relativity is correct.  However, the physical relevance of this 
result is highly questionable -- since it is precisely in such strong
gravity regimes that we should doubt the validity of Einstein's general 
relativity as the correct description of gravity. First of all, for strong
gravitational fields both quantum behaviour of matter fields and 
space-time itself will probably become important and significantly change
the field equations for gravity. The proper understanding of such regimes 
therefore requires a proper knowledge of quantum theory of gravity, which 
is, of course, still currently not available. On the other hand, assuming 
the actual physical existence of singularities would mean the capitulation 
with the respect to the fundamental goal of physics -- namely, the 
complete, non-divergent and consistent description of reality, 
including the evolution of the Universe. For all this reasons, we should 
view the Hawking singularity theorems more as a signal of incompleteness of
Einstein's general theory of relativity, than the proof for the actual 
physical existence of singularities. Furthermore, we believe that the demand
for singularity-free solutions constitutes one of the most important
criteria for the future quantum theory of gravity, and therefore also for
the effective theories which are being investigated in order to overcome
the current gap between the quantum physics and 
description of gravity as a geometry of space-time. 
\\ \\ 
Various investigations in the past decades have demonstrated that even 
simple modifications of gravitational Lagrangian with respect to the
standard Einstein-Hilbert action, while leaving all other physical
assumptions of Einstein's general relativity intact, can prevent the 
appearance of the big-bang singularity \cite{bounce1, bounce2, bounce3, 
bounce4, bounce5,bounce6, bounce7, bounce8, bounce9, bounce10, bounce11, bounce12, effectivefield1, effectivefield2}. Also, in higher curvature gravity theories 
some important 
non-singular investigations have been done in Gauss-Bonnet higher curvature
gravity \cite{Nojiri:2022xdo, Elizalde:2020zcb, Odintsov:2022unp} and
in $f(R)$ gravity theories \cite{Odintsov:2020zct, Odintsov:2020fxb, 
Odintsov:2021yva}.
In such models the big bang singularity is then replaced by the 
cosmological bounce, in which the Universe undergoes a transition from 
contraction to expansion. It is worth remembering that there are in 
principle no real physical reasons for favoring Einstein-Hilbert action 
in comparison to higher curvature modifications, such as $f(R)$, $f(T)$ or 
higher derivative gravity theories, as long as they lead to the same 
observable weak field limit and have no theoretical pathologies. 
As a matter of fact, the Einstein-Hilbert action was historically 
introduced as the simplest action leading to the Newtonian limit, 
out of the infinitely many other equally possible options. 
The investigation of possible modifications of gravitational action, 
and their consequences on the existence of the big bang singularity, are 
therefore important not just for trying to overcome the limitations of 
general relativity - such as the existence of singularities, but also as 
the path for better understanding the structure of potential quantum theory 
of gravity.  \\ \\
There were also numerous works which demonstrated the possibility of the 
cosmological bounce, if some new hypothetical additions to the standard 
cosmological model are added, such as specific scalar fields, extra 
dimensions or new types of fluids \cite{mix1,mix2,mix3,mix4,mix5, mix6, 
mix7, mix8}. Although such investigations can lead to some important 
insights regarding the problem of initial singularity in the cosmological 
evolution, we think that, from a methodological point of view, the 
approach based on the modification of field equations, assuming no new
ingredients, should be viewed as superior. This is because the addition 
of new hypothetical structures and substances  should be disfavored with
respect to explanation which assumes no new unobserved and yet unverified 
forms of matter-energy or spacetime structure. To put it in different 
words, it is always possible to obtain the desired physical goal by 
increasing the number of parameters and invoking various types 
of \textit{ad hoc} entities, but by doing so, the physical theory 
looses its simplicity, necessity and integrity.   
\\ \\ 
One of the important limitations of the usual strategy of investigating 
bouncing cosmologies based on specific constructions (i.e. the specific 
type of modifications of Einstein's general relativity or matter-energy 
content of the Universe) is that the obtained results are highly dependent 
on the specific assumptions which are taken to derive them. It is thus not
easy, and sometimes simply not possible, to see which of the properties of 
solutions are general and which are the result of highly hypothetical and 
often not properly motivated new modifications and additions. This problem 
signifies the need for a model independent study of bouncing cosmologies.
We first tried to contribute to this research program by studying the 
bouncing and cyclic solutions supported by a general type of higher order 
curvature corrections \cite{prviciklicki}. This research was then extended 
and generalized by studying the model independent dynamic properties of 
bouncing cosmologies and then applying the results to different types of 
modified gravity theories \cite{drugiciklicki}. It was also demonstrated 
that the problem of magnetogenesis has a simple possible solution in the 
model-independent approach to bouncing cosmology \cite{magnetogeneza}. 
Recently, we proposed a model-independent approach to bouncing cosmology 
in which we proposed a simple solution of the cosmological constant problem
and also studied the effects of quantum fluctuations of the spacetime 
geometry \cite{quantumgrav}. We will use some of the obtained conclusion 
in the present study of wormholes created during the cosmological bounce.  
\\ \\
There is an obvious technical connection between cosmological bounce and
another type of hypothetical gravitational solutions - wormholes, that comes
from the fact that both types of solutions imply physics beyond standard 
general relativity. Wormholes are solutions of field equations for gravity
which represent a shortcut through spacetime, a tube-like structure which is
asymptotically flat at both ends, and connects two distant parts of the 
Universe. While such spacetime configuration is actually a solution of 
classical Einstein's equations, it requires the violation of Weak Energy 
Condition (WEC) for its existence in the framework of Einstein's gravity 
\cite{morristhorne}. However, it is at the present time not known which kind
of substance could lead to the needed violation of the usual WEC on 
macroscopic scales. Therefore, if WEC is assumed to be satisfied then 
wormholes can be realized only by virtue of modification of field equations
for gravity. Numerous realizations of such wormhole solutions not requiring
the WEC violation were extensively studied in the literature \cite{lobo1, lobo2, lobo3, prvinas, amirabi, habib, yousaf, lobato, godani, zubair, mishra, mustafa,
reza, mandal, ahmed, mubasher, genc, ortiz, Dai:2018vrw, Dai:2020rnc}. One special case of potential
wormhole solutions supported by modified gravity are ``small cosmological
wormholes'', an approximate type of solution where for a large enough 
distance from the wormhole throat the spacetime geometry can be described by
the standard cosmological FLRW metric \cite{kozmocrvi}. Such type of 
solutions enable us to simply study wormholes which are contained in the 
expanding Universe. \\ \\
As we discussed, both the bouncing cosmological solutions and wormholes can
be supported by an effective violation of usual energy conditions coming 
from the additional terms in equations with respect to Einstein's gravity,
playing the role of effective pressures and energy densities, while 
preserving the energy conditions for the matter content of the Universe.
This naturally leads to the question: what is the relationship between the
cosmological bounce and potential existence of wormholes? We address this
question in the present paper where we show that if the conditions for the
cosmological bounce are established, then 
wormholes can exist without any further modification of field equations or 
without introducing any kind of exotic matter. To do this we first present a
simple, very general and model-independent, description of bouncing 
cosmology and then show how cosmological wormholes can be further 
constructed on such spacetime. We then demonstrate that there is no 
violation of WEC in the matter sector implied for such a kind of solutions.
\\ \\
The paper is organized in the following manner: in Section II we discuss how
to generally represent all possible types of cyclic cosmologies on FLRW 
spacetime under the assumption of the stress-energy tensor conservation, in
Section III we discuss the wormhole geometry around the cosmological bounce,
in Section IV we derive field equations for wormholes around the
cosmological bounce, obtain the solutions and discuss the WEC violation, and
we finally conclude in V.

\section{General approach to bouncing cosmology}
In order to study the possibility of wormhole existence during the 
cosmological bounce we first need to present a general (i.e. model 
independent) description of bouncing cosmological solutions. Here we 
follow the steps discussed in \cite{quantumgrav}. With this aim we 
introduce the following action for gravity ($c=1$):
\begin{equation}
    \mathcal{S}_{eff}= \frac{1}{16 \pi G}\int\sqrt{-g} Rd^{4}x + 
    \mathcal{S}_{mod},
    \label{akcija}
\end{equation}
here $\mathcal{S}_{mod}$ is a general type of correction needed to 
support the cosmological bounce. 
We add the action for matter fields and then vary this action 
with respect to the metric, thus obtaining the effective field 
equation for gravity:
\begin{equation}
    G_{\mu \nu}^{eff}= G_{\mu \nu} + G_{\mu \nu}^{mod}= 8 \pi G
    (T_{\mu \nu}^{rad, mat}+ T_{\mu \nu}^{vac}),
    \label{mod}
\end{equation}
here $G_{\mu \nu}$ is Einstein's tensor, $G_{\mu \nu}^{mod}$ simply 
denotes the collection of terms which arise after variation of 
$\mathcal{S}_{mod}$ with respect to the metric, $T_{\mu \nu}^{rad, 
mat}$ is stress-energy tensor for matter and radiation and $T_{\mu 
\nu}^{vac}$, is the stress-energy contribution of the quantum vacuum. 
At this point we make no restrictions on the properties of $G_{\mu 
\nu}^{mod}$, apart from assuming that it is a proper tensor and that 
the total stress-energy tensor is conserved, which leads to 
$\nabla^{\mu}G_{\mu \nu}^{eff}=\nabla^{\mu}G_{\mu \nu}^{mod}=0$. 
We now focus on the spherically symmetric and isotropic type of 
cosmological spacetime, that is we turn our attention on the 
Friedmann–Lema\^{i}tre–Robertson–Walker (FLRW) spacetime, given by the 
following form
\begin{equation}
 ds^{2}=-dt^{2}+a(t)^{2}\bigg( dr^{2}+ r^{2}(d\theta^{2} +  \sin^{2} 
 \theta d\phi^{2})  \bigg).
 \label{flrwmetrika}
 \end{equation}
Although the collection of terms $G_{\mu \nu}^{mod}$ can in principle 
be some arbitrary complicated function of the curvature invariants 
and their combinations (such as Ricci scalar, contractions of Riemann 
and Ricci tensors, their higher powers and derivatives), the only 
cosmological coordinate on which components of this collection on 
terms can depend is - due to the homogeneous and isotropic nature of 
the considered spacetime --  the time coordinate. Therefore, despite 
the arbitrary nature of such corrections, on the FLRW spacetime the 
field equations must take the following form \cite{quantumgrav}:
\begin{equation}
3H^{2}= 8 \pi G \bigg(\rho_{rad}^{0} a^{-4}+ \rho_{mat}^{0}a^{-3} + 
\rho_{vac} \bigg) + S(t) .
\label{modfri1}
\end{equation}
\begin{equation}
    \frac{\ddot{a}}{a}= - \frac{4 \pi G }{3}
    \bigg(\rho_{mat} + \rho_{rad} + 3p_{rad} + \rho_{vac}+3p_{vac} 
    \bigg) + \frac{G_{rr}^{mod}(FLRW)}{2a^{2}}- \frac{S(t)}{6}
    \label{modfri2}
\end{equation}
Here $S(t)=-G_{00}^{mod}(\textit{FLRW})$, 
$T_{00}^{vac}=-\rho_{vac}g_{00}$ and $T_{ij}^{vac}=-p_{vac}g_{ij}$, 
while also using the standard description of matter and radiation 
modelled by the ideal fluid with the equation of state $w=0$ and 
$w=1/3$ respectively. From the demand for the conservation of the 
stress-energy tensor it follows that $\nabla^{\mu}G_{\mu 
\nu}^{mod}=0$, and from that we obtain the following identity for 
$G_{r r}^{mod}$ on FLRW spacetime:
\begin{equation}
\frac{3H}{a^{2}}G_{r r}^{mod}(FLRW)=\dot{S}(t)+ 3H S(t).
\end{equation}
It is simple to check that (\ref{modfri2}) is, under these 
assumptions, just a time derivative of (\ref{modfri1}), and thus we 
will in further analysis inspect mostly equation (\ref{modfri1}). 
Note that (\ref{modfri1}) and (\ref{modfri2}) are the most general 
form of Friedmann equations coming from the purely mathematical modification of the 
action for gravity with respect to general relativity -- limited only by the symmetries of the spacetime 
and the total stress-energy conservation. Of course, when additional physical degrees of freedom are included in the action, such as scalar fields or various forms of non-minimal coupling, the modified equations will not have the form of equations (\ref{modfri1}) and (\ref{modfri2}), but such theories are beyond our interest here. In other words, in this paper we are only concerned with type of theories we could call minimally physically modified with respect to general relativity -- i.e. involving only the mathematical generalizations of the action for gravity and no extra physical degrees of freedom.   \\ \\
Now we need to fix the correction function $S(t)$ in order that it 
leads to a bouncing cosmology. Here also, as we are interested in 
reaching general conclusions regarding the relationship between 
wormholes and cosmological bounce, we do not want to subscribe to 
some specific bouncing scenario, but to remain as general as 
possible. 
For any bouncing solution it necessarily follows that, at the time of 
the bounce, $t=t_{b}$, where the scale factor reaches its minimal 
value, $a=a_{min}$, we have $H=0$ and therefore using (\ref{modfri1}) 
we obtain the following condition: 
\begin{equation}
S(t_{b})=- 8\pi G \bigg(\rho_{rad}^{0} a_{min}^{-4}+ 
\rho_{mat}^{0}a_{min}^{-3} + \rho_{vac} \bigg) .   
\label{sbounce}
\end{equation}
Furthermore, since $t=t_{b}$ represents the minimum of the scale 
factor, around the bounce 
$a(t) \approx a_{min} + (c/2)(t-t_{b})^{2}$, where $c$ is a positive 
constant. From this, one easily obtains the approximation for $H(t)$ 
around the bounce valid for any possible type of bouncing cosmology. 
This approximate form of $S(t)$ around the bounce is given by \cite{quantumgrav}
\begin{equation}
\begin{split}
S(t)_{bounce} \approx 3 \bigg(\frac{c(t-t_{b})}{a_{min}+c (t-
t_b)^{2}/2} \bigg)^{2} - 8 \pi G \Bigg[
\rho_{rad}^{0} \bigg(a_{min} + c(t-t_{b})^{2}/2 \bigg)^{-4} 
\\ \\
+\rho_{mat}^{0} \bigg(a_{min} + c(t-t_{b})^{2}/2 \bigg)^{-3} + 
\rho_{vac} \Bigg ].
\label{saroundbo}
\end{split}
\end{equation}
In order to describe the general solutions for cosmological wormholes 
on bouncing spacetime we need to match the geometry of the 
cosmological spacetime possessing a bounce with a wormhole geometry. 
We will do this in the following section by considering the wormhole 
geometries which asymptotically approach the bouncing cosmological 
solutions. 

\section{The geometry of wormholes around the bounce}

Firstly, we introduce the geometry of a static spherically symmetric 
wormhole \cite{morristhorne}
\begin{equation}
    ds^2=-e^{2\Phi (r)} dt^2 + \frac{dr^2}{1-\frac{b(r)}{r}} +
    r^2(d\theta^2 + \sin^2 \theta d\varphi^2),
    \label{crv}
\end{equation}
where $\Phi(r)$ is sometimes called the redshift function
and $b(r)$ is the shape function, respecting the condition $1-b(r)/r 
\geq 0$. In order to represent a 
traversable wormhole the redshift function must be 
finite everywhere, therefore there are no horizons.
Also, if (\ref{crv}) is a wormhole geometry then 
the shape function must satisfy the flaring-out
condition \cite{morristhorne}
\begin{equation}
    \frac{b(r) - b'(r)r}{b(r)^2}>0.
\end{equation}
This condition tells us that there exist a throat of a
wormhole which represents the minimal radius of the
geometry given by (\ref{crv}).

In a cosmological setting we are interested in wormholes which
evolve in the cosmological time, whose evolution is 
governed by the scale factor $a(t)$. In that case the
static spherically symmetric wormhole can be generalised
to a spherically symmetric cosmological wormhole \cite{kim}
\begin{equation}
    ds^2 = -e^{2\Phi (r,t)} dt^2 + a(t)^2 
    \Bigg( \frac{dr^2}{1-\frac{b(r)}{r}} +
    r^2(d\theta^2 + \sin^2 \theta d\varphi^2) \Bigg),
    \label{crvcosmo}
\end{equation}
where again the shape function is given by $b(r)$, and the redshift 
is $\Phi(r,t)$ but now it can  also be a function 
of time $t$, and finally $a(t)$ is the scale factor. \\ \\
To asymptotically match such type of the solution to the bouncing 
FLRW cosmology we use the considerations presented in 
\cite{kozmocrvi}. We use the approximation of a small cosmological 
wormhole: we suppose that the spacetime geometry of the wormhole is 
such that the redshift and shape functions are negligible for $r 
>r_{c}$, where the value of $r_{c}$ is chosen in such a manner that 
$r_{c} H <<1$ for all considered times. Under such approximation we 
can treat the wormhole as being effectively confined within the 
region $r<r_{c}$ with no influence on the global geometry of FLRW 
spacetime. In other words, such wormhole can be treated as if its 
asymptotical infinity is actually placed at $r=r_{c}$. We furthermore 
assume that all the dynamical properties of such wormholes are 
determined only by the expansion of the Universe, so that the 
dynamics of the scale factor, $a(t)$, for the wormhole is the same as 
for the Universe. Thus, considering the case of bouncing cosmology, 
the scale factor of the cosmological wormhole appearing in 
(\ref{crvcosmo}) can be computed from modified Friedmann equations 
(\ref{modfri1})-(\ref{modfri2}), taking into account the Taylor 
expansion of the correction factor around the bounce, given by 
(\ref{saroundbo}). \\ \\
The components of the stress-energy tensor will in general not be the 
same in the wormhole region, $r<r_{c}$, and in the cosmological 
region, $r>r_{c}$. In the cosmological region the matter-energy is 
modelled by the ideal fluid with the equation of state 
$p_{cosmo}=w\rho_{cosmo}$, with the EOS parameter $w=0$ for matter, 
$w=1/3$ for radiation and $w=-1$ for quantum vacuum. Here we labeled 
the components of the stress-energy tensor in the cosmological region 
with subscript ``cosmo'' to distinguish them from the components in 
the wormhole region. In the region $r<r_{c}$ matter supporting the 
wormhole will be described by anisotropic fluid, given by
\begin{equation}
    T^{\mu}_{\; \;\nu}=\mbox{diag}\Big(-\rho(r,t), p_r(r,t), p_t(r,t), p_t(r,t)\Big),
    \label{stressmaterije}
\end{equation}
where the tensor components are density, radial pressure and 
tangential pressure respectively. Since the anisotropic fluid supporting 
the wormhole must approach the isotropic fluid of the bouncing 
Universe the following boundary conditions need to be satisfied:
\begin{equation}
    \rho(r_{c},t)=\rho_{cosmo}(t),
\end{equation}
\begin{equation}
    p_r(r_{c},t)=p_{cosmo}(t),
\end{equation}
\begin{equation}
    p_t(r_{c},t)=p_{cosmo}(t),
\end{equation}
also, at $r=r_{c}$ the relationships between the density and 
pressure of the fluid supporting the cosmological wormholes needs to 
go to the cosmological equation of state, for matter, radiation and 
quantum vacuum:
\begin{equation}
p_r(r_{c},t)=p_t(r_{c},t)=w \rho(r_{c},t).
\label{cosmouvjet}
\end{equation}

\section{Wormhole solutions and WEC around the bounce}

We can proceed by deriving the equations of motion for  wormholes
around the bounce. Here we are not interested in supporting the wormhole with some specific type of modification of gravity crafted in such a way so that the WEC for matter fields is satisfied on the wormhole spacetime. We rather want to show that even a minimal modification of gravity, supporting the cosmological bounce, can on its own support the wormhole solutions contained in the FRWL spacetime without the WEC violation. For this reason, while analysing the wormhole solutions, we consider only the type of modification responsible for supporting the cosmological bounce and that is $S(t)$. Of course, in general, modifications of action for gravity will lead to correction functions that will in the wormhole region show also a radial dependence, i.e. $S(t,r)$. However, this radial dependence of modifications if not of interest for the analysis presented here. Adding an extra degree of freedom, coming from the presence of radially dependent correction terms, just makes it additionally easier to achieve the absence of WEC violation, while blurring the question of supporting the wormholes by the existence of the cosmological bounce alone. For this reason, we work in a setting where we neglect any additional effects of the modification of gravity on the wormhole, apart from the correction needed to support the bounce. Physically speaking, this assumption corresponds to considering such theories for which 
modifications with respect to general relativity appear only for very high values of curvature reached around the bounce, while leading to negligible departures with respect to general relativity for the changes on the spatial scales of the wormhole (thus making $S$ practically constant with respect to $r$). \\ \\
Equations of motions (\ref{modfri1}) and (\ref{modfri2})
are valid for the isotropic case, however in our case the fluid becomes 
anisotropic as generally $p_r(r,t) \neq p_t (r,t)$ for the wormhole 
geometry. Therefore the 
modified Einstein's equations are slightly different in the anisotropic case.
Going back to equation (\ref{mod}) we define $G^\mu_{\;\; \nu}{}^{mod}=
-8 \pi G T^\mu_{\;\; \nu}{}^{mod}$ where
\begin{equation}
    T^\mu_{\;\; \nu}{}^{mod}=\mbox{diag}\Big(-\rho^{mod}(t), p_r^{mod}(t),
    p_t^{mod}(t), p_t^{mod}(t)\Big),
\end{equation}
\begin{equation}
    \rho^{mod}(t)=\frac{S(t)}{8 \pi G},
\end{equation}
\begin{equation}
    p_r^{mod}(t)=\frac{-1}{8 \pi G}\Big(\frac{a(t) \dot{S}(t)}{ \dot{a}
    (t)( 3+ r\Phi'(t,r)  )}+S(t)\Big),
\end{equation}
\begin{equation}
    p_t^{mod}(t)=\frac{-1}{8 \pi G}\Big(\frac{ a(t) \dot{S}(t)
    ( 2 + r \Phi'(t,r) )
    }{2 \dot{a}(t)( 3+ r\Phi'(t,r) )}+S(t)\Big),
\end{equation}
where the condition $\nabla^\mu T_{\mu\nu}^{mod}=0$ is satisfied from the
definition above. In this case the field equations are
\begin{equation}
    H_{\mu \nu}=8\pi G ( T_{\mu \nu}^{mod} + T_{\mu \nu}) - G_{\mu \nu}=0.
    \label{cijela}
\end{equation}
By putting the wormhole geometry (\ref{crvcosmo}) in field equation
(\ref{cijela}) we obtain the following equations of motion
\begin{equation}
    H^{t}{\;}_{t}=8 \pi G \rho (t,r)+S(t)-\frac{3 \dot{a}(t)^2 
    e^{-2 \Phi (t,r)}+\frac{b'(r)}{r^2}}{a(t)^2}=0,
    \label{prva}
\end{equation}
\begin{multline}
    H^{r}{\;}_{r}=\frac{a(t) \dot{S}(t)}{ \dot{a}(t)(3 + r\Phi'(t,r)   )}+S(t)- \\
    \frac{r^3 \dot{a}(t)^2 e^{-2 \Phi (t,r)}+2 r b(r) \Phi'(t,r)-2 r^2
   \Phi'(t,r)+b(r)}{r^3 a(t)^2}- \\
    \frac{2 e^{-2 \Phi (t,r)} \left(\ddot{a}(t)-
   \dot{a}(t) \dot{\Phi}(t,r)\right)}{a(t)}-8 \pi G p_r(t,r)=0,
   \label{druga}
\end{multline}
\begin{multline}
   H^{\theta}{\;}_{\theta}= \frac{1}{2 a(t)^2}\Big(-2 \dot{a}(t)^2 e^{-2 \Phi (t,r)}+\\
   \frac{2 r \left(r \Phi'(t,r)^2+\Phi'(t,r)+
    r \Phi''(t,r)\right)-b'(r) \left(r \Phi'(t,r)+1\right)}{r^2}- \\
   \frac{b(r) \left(2 r^2 \Phi'(t,r)^2+2 r^2 
   \Phi''(t,r)+r \Phi'(t,r)-1\right)}{r^3}\Big)       
   +   
   \frac{a(t) \dot{S}(t) \left(r \Phi'(t,r)+2\right)}{2
   \dot{a}(t) \left(r \Phi'(t,r)+3\right)}+\\
   S(t)-\frac{2 e^{-2 \Phi (t,r)} 
   \left(\ddot{a}(t)-\dot{a}(t) \dot{\Phi}(t,r)\right)}{a(t)}-8 \pi G p_t(t,r)=0,
   \label{treca}
\end{multline}
\begin{equation}
    H^{t}{\;}_{r}=-\frac{2\dot{a}(t) e^{-2 \Phi (t,r)} \Phi '(t,r)}{a(t)}=0,
    \label{cetvrta}
\end{equation}
where the dot denotes the time derivative and the prime the derivative with respect to the radial coordinate $r$.
From the off diagonal equation (\ref{cetvrta}) it 
appears that either $\Phi'(t,r)=0$ or $\dot{a}=0$, we 
choose the condition $\Phi'(t,r)=0$ in order to 
support the idea of a cosmological wormhole. Also, 
for simplicity and in order to be sure that the
wormhole is free from horizons we choose
$\Phi(t,r)=0$.

We rescale the parameters in order to have a dimensionless system of 
equations in the following manner
\begin{equation}
    \Omega = \frac{\rho}{\rho_c}, \qquad \tilde{p}=
    \frac{p}{\rho_c}, \qquad \tilde{t}=H_0 t, \qquad \tilde{b}=H_0 b, 
    \qquad \tilde{r}=\frac{r}{r_0},
    \label{reskaliranja}
\end{equation}
where $H_0^2=8\pi G \rho_c/3$ and $r_0$ is the radius of the throat. In
principle $H_0$ and $r_0$ are independent quantities, however, as mentioned 
before the critical case is $r_0=1/H_0$. This point represents the upper 
limit for the dimension of wormholes (the throat radius is 
$r_0 \sim 10^{26}$ m for the current measured $H_0$), and in this regime the
assumption of a small cosmological wormhole, $H_0 r_0 < < 1$, obviously does not hold anymore.
Interestingly, in our calculations there are no significant 
qualitative differences for wormholes with throats smaller or bigger
than $r_0=1/H_0$ with $10^{\pm 10}$ orders of magnitude due to a highly
suppressed radial functions in the field equations with respect to the
scale factor in the early Universe. In other words, as will be discussed 
bellow, for the considered realistic values of parameters the fluid supporting
the wormholes approaches the limit of ideal cosmological fluid very fast away
from the throat. This confirms the robustness of our discussion.

Now we need to inspect the Weak
Energy Condition (WEC) at the throat ($r=r_0$) in order to support a
wormhole. The WEC in our rescaled quantities for the stress-energy tensor 
given by (\ref{stressmaterije}) reads \cite{visser, primer}
\begin{alignat}{1}
\Omega \geq 0 ,  \\
\Omega + \tilde{p}_{radial} \geq 0 ,  \\
\Omega + \tilde{p}_{tangential} \geq 0.
\end{alignat}
By combining equations of motion (\ref{prva})-(\ref{treca}) with the
re-scaled quantities (\ref{reskaliranja})  we can 
deduce the following useful expressions for the WEC analysis
\begin{equation}
    \Omega(r,t)=-\frac{-3 \tilde{r}^2 \dot{\tilde{a}}(t)^2+\tilde r^2 a(t)^2 \tilde S(t)-\tilde b'(r)}{3 \tilde r^2 a(t)^2},
    \label{prvares}
\end{equation}
\begin{multline}
    \Omega(r,t) + \tilde p_{r}(r,t)=\frac{1}{9 \tilde r^3 a(t)^2 \dot{\tilde{a}}(t)}
    \Big( 3 \tilde r \dot{\tilde{a}}(t) \tilde b'(r)-3 \tilde b(r) 
    \dot{\tilde{a}}(t)+\\
    6 \tilde r^3 \dot{\tilde{a}}(t)^3-
    6 \tilde r^3 a(t) \dot{\tilde{a}}(t) \ddot{\tilde{a}}(t)+\tilde r^3 a(t)^3 \dot {\tilde S}(t)\Big),
    \label{drugares}
\end{multline}
\begin{multline}
    \Omega(r,t) + \tilde p_{t}(r,t)=\frac{1}{18 \tilde r^3 a(t)^2 
    \dot{\tilde{a}}(t)}   
    \Big(3 \tilde r \dot{\tilde{a}}(t) \tilde b'(r)+3 \tilde b(r) 
    \dot{\tilde{a}}(t)+\\
    12 \tilde r^3 \dot{\tilde{a}}(t)^3-12 \tilde r^3 a(t) \dot{\tilde{a}}(t) \ddot{\tilde{a}}(t)+2 \tilde r^3 a(t)^3 \dot {\tilde S}(t)\Big),
    \label{trecares}
\end{multline}
where $\tilde S(t)= H_0^{-2}S(t)$, $\dot{\tilde a}=da/d\tilde{t}$, $\ddot{\tilde a}=da^2/d\tilde{t}^2$  and we set $\Phi(r,t)=0$, based on the 
previously mentioned considerations.

In order to inspect the bouncing region in the
cosmological epoch we will use the approximate 
form of $S(t)$ given by equation (\ref{saroundbo})
$S(t)=S(t)_{bounce}$ where we fixed $a(t)=a_{min} + 
c(t-t_b)^2/2$. In that case we can proceed 
with an example of a wormhole, where we choose the shape function as
\begin{equation}
    b(r)=\frac{r_0^3}{r^2},
\end{equation}
which is one of the simplest function to preserve the flaring-out 
condition and one of the most common form used in literature \cite{kozmocrvi, mimetic}.

\begin{figure}[H]
    \centering
    \includegraphics[scale=0.67]{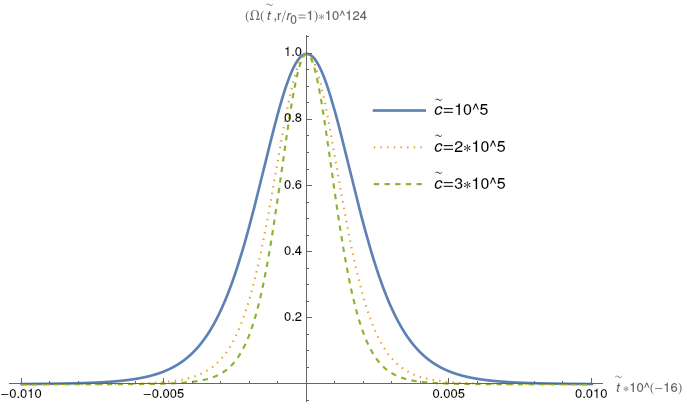}
    \caption{Here we present the density parameter, $\Omega$ divided by factor $10^{124}$, at the
    throat, $r=r_0=10^{-2}/H_0$, for different values of the parameter $\tilde c$.
    We also assumed $\Phi(r,t)=0$, $b(r)=r_0^3/r^2$, $S(t)=S(t)_{bounce}$, $a_{min}=10^{-32}$
    and the cosmological parameters $\Omega^0_{vac}=0.692$, 
    $\Omega^0_{mat}=0.308$ and $\Omega^0_{rad}=10^{-4}$.}
    \label{plot1}
\end{figure}

\begin{figure}[H]
    \centering
    \includegraphics[scale=0.68]{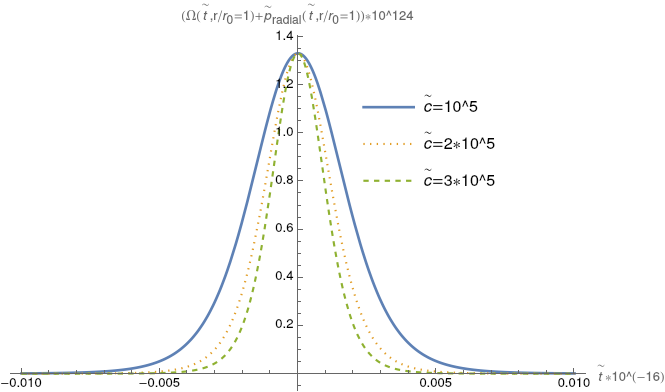}
    \caption{Here we present the component $\Omega + p_r$, divided by factor $10^{124}$, at the
    throat, $r=r_0=10^{-2}/H_0$, for different values of the parameter $\tilde c$.
    We also assumed $\Phi(r,t)=0$, $b(r)=r_0^3/r^2$, $S(t)=S(t)_{bounce}$, $a_{min}=10^{-32}$
    and the cosmological parameters $\Omega^0_{vac}=0.692$, 
    $\Omega^0_{mat}=0.308$ and $\Omega^0_{rad}=10^{-4}$.}
    \label{plot2}
\end{figure}

\begin{figure}[H]
    \centering
    \includegraphics[scale=0.62]{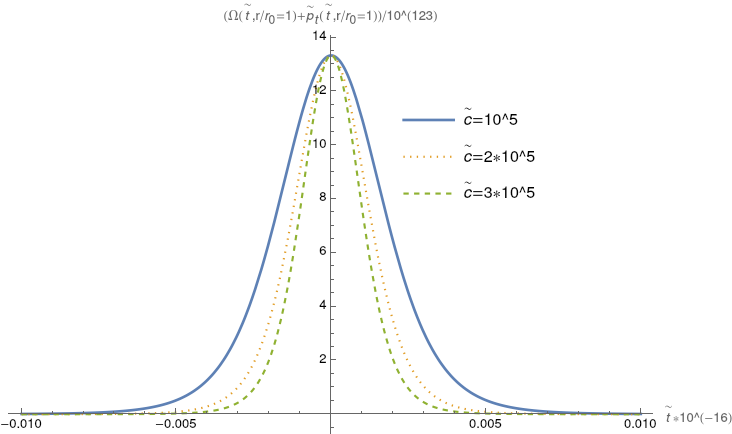}
    \caption{Here we present the component $\Omega + p_t$, divided by factor $10^{123}$, at the
    throat, $r=r_0=10^{-2}/H_0$, for different values of the parameter $\tilde c$.
    We also assumed $\Phi(r,t)=0$, $b(r)=r_0^3/r^2$, $S(t)=S(t)_{bounce}$, $a_{min}=10^{-32}$
    and the cosmological parameters $\Omega^0_{vac}=0.692$, 
    $\Omega^0_{mat}=0.308$ and $\Omega^0_{rad}=10^{-4}$.}
    \label{plot3}
\end{figure}

Firstly, we present the time evolution of the throat ($r=r_0$) around the
cosmological bounce for the density parameter $\Omega(r=r_0,t)$ on Fig. \ref{plot1}
The other two conditions for the  WEC are depicted in Fig. \ref{plot2} 
and Fig. \ref{plot3}.
It can be seen that at the cosmological bounce the WEC is ``maximally''
satisfied at the throat in the sense that the inequalities for the 
WEC take maximally positive values. 
The value of the parameter $c$ dictates how further
in time will the WEC be satisfied. Namely, with the higher value of $c$ the scale
factor more rapidly increases, leading to a higher acceleration of the
Universe. This claim can also be supported by a simple analysis of equation
(\ref{drugares}).   
Firstly, the equation (\ref{saroundbo}) can be rewritten in a more general (and dimensionless) form
\begin{equation}
   H_0^{-2}S(t)_{bounce}= \tilde{S}(t)_{bounce}=3\Big(\frac{\dot{\tilde a}}{a}  \Big)^2 - 3\Big( 
    \Omega^0_{rad}a^{-4} + \Omega^0_{mat}a^{-3} - \Omega_{vac}\Big),
\end{equation}
by taking the time derivative, $d/d\tilde{t}$, of the above 
equation we obtain
\begin{equation}
    \dot{\tilde{S}}(t)_{bounce}=\frac{3 \dot{\tilde{a}}}{a^5}(  4\Omega^0_{rad} +3 \Omega^0_{mat}a - 2a^2 \dot{\tilde{a}}^2+ 2a^3 \ddot{\tilde{a}}).
    \label{tildanistocka}
\end{equation}
Finally, by using $\dot{\tilde{S}}(t)_{bounce}$ from equation 
(\ref{tildanistocka}) and putting it in equation (\ref{drugares}) we get 
the following, remarkably simple, result:
\begin{equation}
    \Omega(r,t) + \tilde p_r(r,t)=\frac{4}{3}\frac{\Omega^0_{rad}}{a^4} + 
    \frac{\Omega^0_{mat}}{a^3}-\frac{\tilde b-\tilde r \tilde b'}{3\tilde r^3 a^2}.
    \label{prvawec}
\end{equation}
The first two positive contributions decrease with the power of $a^4$ 
and $a^3$ but the last negative contribution ($b-r b'$, which is 
nothing else than the flaring-out condition) decrease with the power of
$a^{2}$. This explains why at the bounce where $a$ is minimal the WEC
is satisfied. Similarly, but in a bit more relaxed way, the third
condition (\ref{trecares}) behaves like
\begin{equation}
    \Omega(r,t) + \tilde p_t(r,t)=\frac{4}{3}\frac{\Omega^0_{rad}}{a^4}
    + \frac{\Omega^0_{mat}}{a^3} +\frac{4}{9}\Big( \frac{H}{H_0}\Big)^2 +
    \frac{\tilde b+\tilde r\tilde b'}{6a^2} - \frac{4}{9}\frac{\ddot{\tilde{a}}}{a},
\end{equation}
where the positive acceleration $\ddot a$ dominates at late times.

In our model for $S(t)=S(t)_{bounce}$ it is possible to find at which
times the WEC is broken. The most difficult condition to satisfy is the condition given by equation (\ref{prvawec}), by solving it with
$a(t)=a_{min}+ct^2/2$, ($t_b=0$) we get  four solutions
\begin{equation}
    \tilde t=\pm \Bigg( \frac{2 a_{min} \tilde c( \tilde r \tilde b'- \tilde b) \pm \sqrt{\tilde c^2 \tilde r^3 \Big( 16\Omega^0_{rad} (\tilde b- \tilde r  \tilde b') +9 \tilde r^3 (\Omega^0_{mat})^2\Big)}+3 \tilde c \tilde r^3   \Omega^0_{mat}}{\tilde c^2 \left(\tilde b-\tilde r \tilde b' \right)}\Bigg)^{1/2},
\end{equation}
from which the real solutions, in the case $b(r)=r_0^3/r^2$, 
$\tilde c=10^5$ with the 
cosmological parameters $\Omega^0_{rad}=10^{-4}$ and 
$\Omega^0_{mat}=0.308$, results in
$t\sim \pm 0.00248/H_0$. Again, the result highly depends on the parameter
$\tilde c$, on the acceleration of $a$ around the bounce, where for example 
with $\tilde c\sim 0.617$ the time approaches to $1/H_0$. However, we should remember that such calculations can be assumed to be valid only around the bounce, so high values of the time interval for the existence of wormholes, comparable to the age of the Universe, obviously have no physical interest -- since after the bounce is finished the functional forms for $a(t)$ and $S(t)$ assumed here are no longer valid. We can thus conclude from this discussion that for proper values of the acceleration of $a(t)$, cosmological wormholes can exist as long as the bouncing phase. Interestingly, the
solutions do not depend on the constant vacuum energy $\Omega_{vac}^0$ as
 only the derivative $\dot{S}(t)$ appears in the equation (\ref{drugares}).
\\
\\
We can also analyze the asymptotic, $r \rightarrow \infty$, behaviour of 
the wormhole. Firstly, it can be seen that from the flaring-out condition
$b(r)-b'(r)r>0$ the critical asymptotic case  ($b(r)-b'(r)r=0$) is given by
\begin{equation}
    \lim_{r \to +\infty} b(r) \sim r,
\end{equation}
therefore, functions $b(r)$ which have the asymptotic behaviour with 
greater power in $r$ are not allowed as a representation of a wormhole. 
By taking this limit in the field equations (\ref{prva})-(\ref{cetvrta}) 
we indeed recover the 
cosmological equations given in \cite{quantumgrav} with the properties 
given by (\ref{cosmouvjet}) when $r \rightarrow \infty$. Therefore, the fluid supporting the 
wormhole for large $r$ approaches the values of pressures and densities 
of the cosmological fluid fast, thus confirming the consistency of the
approximation of a small cosmological wormhole set up in a FLRW 
spacetime  and supported by the cosmological bounce. It can also be easily seen that equations are everywhere regular for all the values of $r$ at and away from the throat, and that they approach the homogeneous and isotropic limit fast for $r>r_{0}$. Since the plots for $r$ dependence of the fluid densities and pressures thus do not reveal any new information, while WEC is simply satisfied away from the throat, and the differences in values with respect to the cosmological fluid values are minimal, we do not show those plots here.

\section{Discussion and conclusion}
In this work we have discussed the existence of wormholes during the 
cosmological bounce without the need for additional exotic matter supporting
the wormhole. We have considered the approximation of a  small cosmological
wormhole, whose fluid components needed for supporting its geometry, for
large enough values of $r$, approach the components of the ideal 
cosmological fluid and whose dynamics is dictated by the time evolution of 
the cosmological scale factor, $a(t)$. Thus, this geometry can be treated as
a dynamical wormhole situated in the FLRW spacetime. By considering the most
general form of modified gravity on FLRW spacetime coming from the mathematical generalizations of action, needed for the existence
of bounce replacing the big bang singularity, we have obtained the 
cosmological wormhole solutions which do not need any additional exotic 
matter, that is, which satisfy WEC. 
This is possible since the same modification of gravity which
supports the bounce can at the same time support the existence of
wormholes, and no additional exotic matter is thus needed. The effect leading to simultaneous support of a bounce and a small wormhole geometry is described by the function $S(t)$, and is arising from the generalization of action for gravity, while all the matter fields, $\Omega$, are completely standard and do not contain any new exotic component.
We have thus demonstrated that the 
cosmological bounce in general represents an ideal environment for the natural
creation of wormholes. Such wormholes could then serve as tunnels through
spacetime in the early Universe, connecting very distant points that would
otherwise be causally disconnected. Therefore, the existence of wormholes
around the bounce can further help to solve the horizon problem, relaxing
the necessary duration of the bouncing phase in order to obtain the causal
picture implied by the CMB. We have also demonstrated that such wormholes 
spontaneously start to violate WEC after a certain time, which strongly 
depends on the rate of acceleration of the cosmological scale factor during
the bounce. This can mean that after such critical time wormholes can no 
longer be supported in the Universe and vanish. This opens an interesting 
problem of detailed physical description of ``death of a dynamic wormhole'',
which is left for future work. Solving this problem could also be helpful 
for our better understanding  of how to construct wormholes artificially
(which is just the inverse of the mentioned problem in time).

\end{document}